\documentclass[onecolumn,superscriptaddress,secnumarabic,amssymb,amsmath,nobibnotes,aps,prd,nofootinbib,altaffilletter]{revtex4}
\usepackage{amsmath,amssymb}
\usepackage{graphicx}
\usepackage{dcolumn}
\usepackage{bm,color}
\usepackage{hyperref}
\usepackage{accents}
\usepackage{amssymb,float}
\usepackage{amsmath}
\usepackage{multirow}
\usepackage{siunitx}
\usepackage{tabularx}
\usepackage{booktabs}
\usepackage{url}
\usepackage{orcidlink}

\usepackage{capt-of} 
\usepackage{placeins} 
\usepackage{psfrag}
\usepackage{times}
\usepackage[varg]{txfonts}
\usepackage{xcolor}
\usepackage{xspace}

\makeatletter
\long\def\dddddot#1{%
  {\mathop {#1}\limits ^{\vbox to-1.4\ex@ {\kern -\tw@ \ex@ \hbox {\normalfont .....}\vss }}}%
}
\long\def\multidots#1#2{%
  \count@=0
  {{\mathop {#2}\limits ^{\vbox to-1.4\ex@ {\kern -\tw@ \ex@ \hbox {\normalfont %
  \loop%
  \ifnum#1>\count@%
  .%
  \advance\count@ by1%
  \repeat%
  }\vss }}}}%
}
\makeatother

\newcommand{\ham}{\mathcal{H}}
\newcommand{\lag}{\mathcal{L}}

\begin{document}

\title{Fab-Four cosmography to tackle the Hubble tension}

\author{Celia Escamilla-Rivera\orcidlink{0000-0002-8929-250X}}
\email{celia.escamilla@nucleares.unam.mx}
\affiliation{Instituto de Ciencias Nucleares, Universidad Nacional Aut\'onoma de M\'exico, 
Circuito Exterior C.U., A.P. 70-543, M\'exico D.F. 04510, M\'exico.}

\author{Jos\'e Mar\'ia de Albornoz-Caratozzolo}
\email{jose.albornoz@correo.uia.mx}
\affiliation{Depto. de Física y Matemáticas, Universidad Iberoamericana Ciudad de México, Prolongación Paseo
de la Reforma 880, México D.F. 01219, México.}

\author{Sebasti\'an N\'ajera\orcidlink{0000-0001-9086-343X}}
\email{sebastian.najera@correo.nucleares.unam.mx}
\affiliation{Instituto de Ciencias Nucleares, Universidad Nacional Aut\'onoma de M\'exico, 
Circuito Exterior C.U., A.P. 70-543, M\'exico D.F. 04510, M\'exico.}

%%%%%%%%%%%%%%%%%%%%%%%%%%%%%%%%%%%%%%%%%%%%
%%%%%%%%%%%%%%%%%%%%%%%%%%%%%%%%%%%%%%%%%%%%

\begin{abstract}
In the context of the Fab-Four theory of gravity in a Friedmann-Lemaître-Robertson-Walker background, in this work we use the cosmography approach to study a particular self-tuning filter solution focused on a zero-curvature fixed point to study the $H_0$ tension. In this scheme, the equations restrict the universe's evolution to certain scenarios, including radiation-like expansion, matter-like expansion, and late-time acceleration. Furthermore, we build the cosmographic series of the Fab-Four theory to obtain the kinematic parameters as the Hubble constant $H_0$ and the deceleration parameter $q_0$ for all the scenarios mentioned. Finally, we compare our results to find that it is possible to alleviate the current discrepancy on $H_0$ by considering specific requirements on the free parameters of the Fab-Four theory through a self-tuning filter.
\end{abstract}

\maketitle

%%%%%%%%%%%%%%%%%%%%%%%%%%%%%%%%%%%%%%%%%%%%
%%%%%%%%%%%%%%%%%%%%%%%%%%%%%%%%%%%%%%%%%%%%
\section{Introduction}

The concordance model of cosmology, $\Lambda$~ Cold Dark Matter ($\Lambda$CDM), is thought to work well at explaining several observations of the Universe, including the observed late cosmic acceleration \cite{riess1998observational, 1999Perlmutter}, the structure of the cosmic microwave background (CMB), the baryon acoustic oscillations (BAOs), among others \cite{dodelson_schmidt_2021}. In recent years there have been several issues at hand \cite{Perivolaropoulos_2022}, e.g. CMB anisotropy anomalies \cite{Schwarz_2016}, cosmic dipoles \cite{Perivolaropoulos_2022}, and the most notable issue: the Hubble constant tension, reaching a statistical discrepancy of $\sim$5$\sigma$ \cite{Planck_2020, kamionkowski2022hubble, Riess_2022, DiValentino:2020zio} between early and late measurements.

Direct measurements of the current Hubble constant $H_0$, through late-universe observations, have yielded a value of $H_0 = 73.3 \pm 1.04$ km s$^{-1}$ Mpc $^{-1}$\cite{Riess_2022}, while early-universe observations of the CMB give a lower result of $H_0 = 67.4 \pm 0.5$ km s$^{-1}$ Mpc $^{-1}$ \cite{Planck_2020} differing by around 8-10\%. This discrepancy has positioned itself as an important crisis for the standard $\Lambda$CDM model. Consequently, several theories have been proposed to solve, alleviate or give insights into the Hubble tension issue. In the same line of thought, some proposals focus on the nature of dark energy, since if the cosmological constant $\Lambda$ played a bigger role than previously thought in the early stages of the expansion of the Universe, it could alter the calculation for the early-universe Hubble constant \cite{Karwal_2016}, as a consequence, this would bring it closer to the locally measured $H_0$ value. On one hand, changing the equation-of-state of dark energy by making it time-dependent, has also been proposed as a possible explanation \cite{Zhao_2017}. Other theories include adding new interactions, such as interacting dark energy \cite{Aich:2022idb,Kim:2023oxg},  interacting dark matter \cite{Cao_2011} and interacting holographic dark energy \cite{Escamilla-Rivera:2022baz}; including sterile neutrinos \cite{Salvio:2021puw}, axions \cite{Berlin:2016bdv}, or other theorized particles to increase the relativistic degrees of freedom; unifying dark energy and dark matter as a single fluid \cite{Farnes:2017gbf}; introducing cosmological piecewise functions \cite{Sandoval-Orozco:2022not}; using statistical lookback time approach \cite{Escamilla-Rivera:2023msg}, considering the black hole shadows as standard rulers calibrators \cite{Escamilla-Rivera:2022mkc}, among many others \cite{Di_Valentino_2021, Adhikari:2022sve,Abdalla:2022yfr}. See for example \cite{Schoneberg:2021qvd} for a summary and comparison of several proposals that involve interesting solutions to the $H_0$ tension.

On the other hand, alternative gravity theories also attempt to solve the Hubble tension and other problems within standard cosmology. These theories change General Relativity's (GR) Einstein-Hilbert action to introduce different dynamics in the resulting equations of motion, which can explain different phases of expansion of the universe and could explain the Hubble tension \cite{Kobayashi_2019, Clifton:2011jh}. Allowing higher order equations of motion, there are theories like $f(\mathcal{R})$ theories, which introduce a function of the Ricci scalar, $\mathcal R$, to act as the action. In the context of Teleparallel Gravity \cite{Bahamonde:2021gfp} it can be considered $f(\mathcal{T})$ theories, which instead use functions of the torsion scalar. Another way to extend GR is by allowing the Lagrangian density to depend not only on the metric but also on another degree of freedom such as a scalar field $\phi$, these are the so-called scalar-tensor theories. If we maintain the requirement of working in four dimensions, and having second-order equations of motion -- to avoid Ostrogradsky instabilities \cite{Kobayashi_2019}-- the most general of these theories is Horndeski theory \cite{Charmousis_2012,Horndeski:1974wa}.

In this work, we focus on a special case of Horndeski's theory first proposed in \cite{Charmousis_2012} called the \textit{Fab Four} theory. As a first approach, in \cite{Charmousis_2012} a \textit{self-tuning} filter was applied on the original Horndeski action that partially de-couples the curvature from the cosmological constant and reduces the action to only four base Lagrangians, each one a function dependent on combinations of the Riemann tensor and an arbitrary function of the scalar field $\phi$. The self-tuning filter consisted in requiring that \cite{Charmousis_2012}:
\begin{enumerate}
    \item The theory admits a patch of Minkowski vacuum for any value of the cosmological constant $\Lambda$, and for instantaneous changes in its value;
    \item The theory allows non-trivial cosmologies by introducing additional dynamical terms to the standard one.
\end{enumerate}
According to these couple of requirements, in \cite{Copeland:2012qf} an initial outlook was provided on the types of cosmology and its dynamics derived from the Fab-Four theory. Although it is a fairly recent proposal, the Fab-Four theory has been explored extensively and in different contexts. Specifically, there have been several studies involving the $\lag_{john}$ term in isolation, since it is the one that mainly drives the self-tuning mechanism \cite{Latosh:2018xai} and it contains a derivative coupling to the scalar field, e.g. in \cite{Linder:2013zoa} the $\lag_{john}$ term (see Sec.~\ref{sec:fabfour-intro}) is studied in detail, specifically its generalization as the $\lag_5$ term in Horndeski's theory which adds a dependency on the field kinetic term $X=\nabla_\mu \phi \nabla^\mu \phi$, finding that the resulting \textit{Fab 5} theory can be ghost-free, or stable, or self-tuning, but not all these characteristics at the same time. \textit{Fab 5} cosmology is studied in more detail in \cite{Starobinsky_2016, Starobinsky_2020, Galeev_2021}. After the extension of the original Horndeski Lagrangian \cite{Gleyzes_2015}, adding two more terms in Horndeski's action, it was shown in \cite{Babichev:2015qma} that these new terms can be included in a generalized Fab-Four theory where the arbitrary functions of the field $V_j(\phi)$ and $V_p(\phi)$ become arbitrary functions of the field and its kinetic term $V_j(\phi, X)$ and $V_p(\phi, X)$. Additionally, it was shown that these new terms also present self-tuning for screening the cosmological constant. In \cite{Ezquiaga_2017}, it was argued that the measurement of gravitational waves (GWs) speed strongly constrained \textit{John}-like terms in Horndeski's action, along with several other scalar-tensor theories and dark energy (DE) models. 
However, in \cite{Latosh:2018xai} it is shown that working in an Effective Field Theory (EFT) framework, the resulting effective action is not constrained as such, and so it still admits \textit{John}-like terms, therefore allowing a self-tuning mechanism in the EFT regime and the possibility to be consistent with the observational data \cite{Copeland:2018yuh,Ezquiaga:2017ekz}. In order to explore GW constraints, see \cite{Bordin_2020}. Furthermore, as any alternative theory of gravity, Fab Four proposal could be excluded if we demand that the speed of gravitational waves $c_s$ is exactly the same as the speed of light $c$ along the evolution of the Universe. Therefore, if we consider small deviations from $c$, then several models can become consistent with the observable data (and its constrictions).
New self-tuning models known as \textit{well-tempered} have been developed and deal with the GW constraints by construction \cite{Appleby_2018}, and also prevent the scalar field from screening not only the cosmological constant but also all other matter sources in the system.   A generalized class of \textit{well-tempered} models and their self-tuning mechanism has been discussed in \cite{Emond:2018fvv}. In a quantum gravity context, the study of the quantization of the Fab-Four \textit{John} term results in a Hamiltonian with fractional powers of the momenta, leading to the use of a conformal fractional derivative approach to get a wide range of solutions for the scale factor and the scalar field \cite{Torres:2020xkw}.

In this work, we follow closely the derivations performed for the Fab-Four cosmology \cite{Copeland:2012qf} to obtain the corresponding equations of motion for a Friedmann-Lemaître-Robertson-Walker (FLRW) background and to apply a dynamical systems approach to solving the dynamical equations. Under this scheme and considering all four components of Fab-Four theory, we also extend this analysis by using a cosmographic approach, which will be useful to obtain kinematic quantities, such as $H_0$ and $q_0$. Finally, with these quantities at hand, we will be capable to set general requirements on the free parameters to relax the Hubble tension and give a likelihood landscape for future works and analyses of this theory.
The core idea behind this process is the facility to look for a stable cosmological solution through dynamical systems. The advantage of this(these) kind of solution(s) is the possibility to explore the ones that fulfill the cosmological constraint equation of interest for the scheme to analyze. Once this constraint is fulfilled the \textit{analytical} solution at hand can be tested with observations in order to set a cutoff in the cosmological parameters of interest, in our case, $H_0$ and $q_0$. Of course, there could be more analytical solutions for the set of equations, however, since we are considering a cosmographic approach up to second order, this helps us to reduce the number of them and select the ones that fix the late-time cosmic acceleration. 
  
This paper is divided as follows: 
In Sec.~\ref{sec:fabfour-intro}, we present a brief introduction of the results discussed in \cite{Copeland:2012qf}, where the full equations of motion of the system are given for the Fab-Four theory in a FLRW background. Also, we describe the dynamical system approach for this theory to find the dynamical parameters and the corresponding fixed points. We rewrite the equations in the zero-curvature fixed point of interest and solve them to find viable cosmological scenarios. 
In Sec.~\ref{sec:fabfour_cosmography} we construct the cosmographic series for the Fab-Four theory and find the expressions for the Hubble constant $H_0$, and the deceleration parameter $q_0$ for each cosmological scenario described. Also, we will restrict our analysis to a couple of cases due to the solutions found in the dynamical system approach. 
In Sec.~\ref{sec:fabfour_tension} we discuss how our solutions from the Fab-Four theory are capable to relax the $H_0$ tension by following a set of requirements on the free parameters of the theory. Finally, we present our conclusions and considerations for future work in Sec.~\ref{sec:conclusions}.

%%%%%%%%%%%%%%%%%%%%%%%%%%%%%%%%%%%%%%%%%%%%
%%%%%%%%%%%%%%%%%%%%%%%%%%%%%%%%%%%%%%%%%%%%

\section{Fab-Four cosmology as a dynamical system}
\label{sec:fabfour-intro}

The Fab-Four model comes from a restriction set upon Horndeski's theory, which simplifies it greatly so that the Lagrangian density is now just the sum of four base Lagrangians \cite{Charmousis_2012}:
\begin{equation}
\lag_{\text{Fab-Four}}   = \lag_{\text{john}} + \lag_{\text{paul}} + \lag_{\text{george}} + \lag_{\text{ringo}},
\end{equation}
where each Lagrangian is defined as a combination of derivatives of the scalar field $\phi$, combinations of the metric tensor, and an arbitrary function of the field $V_i(\phi)$, with $i=j,p,g,r$, namely 
\begin{eqnarray}
\lag_{john}&=&\sqrt{-g}V_{john}G^{\mu\nu}\nabla_\mu \phi \nabla_\nu \phi,\\
\lag_{paul}&=&\sqrt{-g}V_{paul}P^{\mu\nu\alpha\beta}\nabla_\mu \phi \nabla_\alpha \phi \nabla_\nu \phi\nabla_\beta \phi,\\
\lag_{george}&=&\sqrt{-g}V_{george}R,\\
\lag_{ringo}&=&\sqrt{-g}V_{ringo}\hat{\mathcal{G}},
\end{eqnarray}
where $G_{\mu\nu}$ is the Einstein tensor, $P_{\mu\nu\alpha\beta}$ is the double dual of the Riemann tensor, $R$ is the Ricci scalar and $\hat{\mathcal{G}}$ is the gauss-Bonnet combination \cite{Copeland:2012qf}. 
Given this notation and considering a FLRW metric whose line element is given by
\begin{equation}
    ds^2= -dt^2 + a^2(t)\left(\frac{dr^2}{1-kr^2}+r^2d\Omega^2
\right),
    \label{eq:FLRW}
\end{equation}
where $a(t)$ is the scale factor, which denotes the relative size of the co-moving, constant-time hypersurfaces of space-time, $d\Omega^2 = d\theta^2 +\sin^2 \theta d\phi^2$ is the element of solid angle and $k$ is the spatial curvature constant. Here, $k=0$ gives a flat space, while $k>0$($k<0$) corresponds to positive (negative) curvature. We can express the Fab-Four Lagrangians as \cite{Copeland:2012qf}
\begin{eqnarray}
\lag_{j}&=&3a^3V_{j}(\phi) \dot\phi^2 \left(H^2+ \frac{k}{a^2}\right)
\label{eq:lagjconmetrica},\\
\lag_{p}&=&-3a^3V_{p}(\phi) \dot\phi^3H \left(H^2+ \frac{k}{a^2}\right)
\label{eq:lagpconmetrica},\\
\lag_{g}&=&-6a^3V_{g}(\phi) \left(H^2- \frac{k}{a^2}\right)-6a^3V_{{g},\phi}(\phi) H\dot\phi
\label{eq:laggconmetrica},\\
\lag_{r}&=&-8a^3V_{{r},\phi}(\phi)\dot\phi H \left(H^2+3 \frac{k}{a^2}\right).
\label{eq:lagrconmetrica}
\end{eqnarray}
Here, $V_{i,\phi}$ denotes the derivative of $V_i$ with respect to the field, ${d V_i}/{d \phi}$, and $\dot \phi=U^a\nabla_a \phi$, with $U^a=\tensor{\delta}{a}{\;\;  t}$ the $4$-velocity.
Following \cite{Copeland:2012qf}, we find that in the presence of standard matter and a cosmological constant $\Lambda$, the Hamiltonian density is given by
\begin{equation}
    \ham = \ham_j + \ham_p + \ham_g + \ham_r + \rho_{\Lambda}=0,
    \label{eq:hambuena}
\end{equation}
where, for convenience, we can express each Hamiltonian in conformal time by considering $N=\ln{a}$, where $A'$ denotes differentiation with respect to $N$, such that $A' = {dA}/{dN} = \dot{A}(1/H)$:
\begin{eqnarray}
    \ham_{j}&=&3V_{j}{\phi'}^2 H^2 \left(3H^2+\frac{k}{a^2}\right) \label{eq:hamjN},\\
    \ham_{p}&=&-3V_{p}{\phi'}^3H^4\left(5H^2+3\frac{k}{a^2}\right)
    \label{eq:hampN},\\
    \ham_{g}&=&-6V_{g}\left[\left(H^2+\frac{k}{a^2}\right)+H^2  \frac{V_{{g},\phi}\phi' }{V_{g}}\right]
    \label{eq:hamgN},\\
    \ham_{r}&=&-24V_{{r},\phi}\phi' H^2 \left(H^2+\frac{k}{a^2}\right)
    \label{eq:hamrN}.
\end{eqnarray}
The equations for the scalar field $\phi$ can be obtained through varying
\begin{equation}
    E_i^{(\phi)}=-\frac{d}{dt}\left(\frac{\partial \lag_i}{\partial \dot{\phi}}\right) + \frac{\partial \lag_i}{\partial \phi},
    \label{eq:eulerlagrange}
\end{equation}
which give the following expressions \cite{Copeland:2012qf}:
\begin{eqnarray}
    E^{(\phi)}_j \dot \phi H^{-1}&=&\frac{1}{a^3 \Delta_2} \left(\frac{\ham_j a^6 \Delta_2^2}{3 \Delta_2-2\frac{k}{a^2}} \right)'
    \label{eq:eomj},\\
    E^{(\phi)}_p\dot \phi H^{-1}&=&\frac{2}{a^{3/2} (H \Delta_2)^{1/2}} \left(\frac{\ham_p a^{9/2} (H\Delta_2)^{1/2} \Delta_2 }{5\Delta_2-2\frac{k}{a^2}} \right)'
    \label{eq:eomp},\\
    E^{(\phi)}_g\dot \phi H^{-1}&=&-3\frac{V_g'}{a} (\Delta_2 a^4)'
    \label{eq:eomg},\\
    E^{(\phi)}_r \dot \phi H^{-1}&=& \frac{\ham_r}{H \Delta_2} \left[a^3\left(\frac{k}{a^2}\Delta_1 +\frac{1}{3}  \Delta_3 \right) \right]'
    \label{eq:eomr},
\end{eqnarray}
with $  \Delta_n = H^n - (\sqrt{-k} / a)^n$.
Notice that this function goes to zero as the latter equations approach a self-tuning solution corresponding to a Ricci flat scenario and 
the full system of equations that arises is coupled with second-order differential equations which are extremely difficult to solve analytically. 
Therefore, a viable path to reach a set of possible solutions is through a dynamical systems approach\cite{Copeland:2012qf}.
In order to implement
this procedure, we consider the following dynamical variables:
\begin{equation}
    h=\frac{H'}{H}, \qquad \qquad \sigma=\frac{\sqrt{-k}}{Ha},
    \label{eq:hysigma}
\end{equation}
notice that if $h=$const. we obtain that
$    H=H_0 a^h  \Rightarrow  a \propto t^{-\frac{1}{h}},$
where a matter-like expansion would arise from $h=-3/2$, radiation-like expansion from $h=-2$, and an inflationary expansion coming from $-1<h<0$.
To build the dynamical system according to the latter variables, we define the following expressions \cite{Copeland:2012qf}:

\begin{equation}
    x=H^\alpha \phi',
    \label{eq:x}
\end{equation}
\begin{align}
    &&y_j=H^{4-2\alpha}V_j, 
    &&y_p=H^{6-3\alpha}V_p,
    &&y_g=H^{2}V_g, 
    &&y_r=H^{4-\alpha}V_{r,\phi}, 
    \label{eq:ys}\\
    &&\lambda_j=H^{-\alpha}\frac{V_{j,\phi}}{V_j}, 
    &&\lambda_p=H^{-\alpha}\frac{V_{p,\phi}}{V_p}, 
    &&\lambda_g=H^{-\alpha}\frac{V_{g,\phi}}{V_g}, 
    &&\lambda_r=H^{-\alpha}\frac{V_{r,\phi\phi}}{V_{r,\phi}}, 
    \label{eq:lambdas}\\
    &&\mu_j=\frac{V_j V_{j,\phi\phi}}{(V_{j,\phi})^2},
    &&\mu_p=\frac{V_p V_{p,\phi\phi}}{(V_{p,\phi})^2}, 
    &&\mu_g=\frac{V_g V_{g,\phi\phi}}{(V_{g,\phi})^2}, 
    &&\mu_r=\frac{V_{r,\phi} V_{r,\phi\phi\phi}}{(V_{r,\phi\phi})^2},
    \label{eq:mus}
\end{align}
where $\alpha$ is an arbitrary constant that provides freedom in re-defining the scalar field while maintaining equations of motion that do not explicitly depend on $H$. This freedom will later be fixed to a particular value in Sec.~\ref{sec:fabfour_cosmography}.

The evolution equations for these variables are directly obtained by differentiating Eqs.~(\ref{eq:hysigma})-(\ref{eq:mus}) with respect to $N$. Finally, we obtain
\begin{eqnarray}
    \sigma' &=& \sqrt{-k}\frac{1}{H^2a^2}\left(H' a + Ha\right)=-\sigma(h+1),
    \label{eq:difsigma}\\
    x' &=& \alpha H^{\alpha-1}H'\phi' + H^\alpha \phi'' = x\left(\alpha h + \frac{\phi''}{\phi'}\right),
    \label{eq:difx}\\
    y_i' &=& \beta_i H^{\beta_i}hV_i + H^{\beta_i}V_{i,\phi}\phi' = y_i\left(\beta_i h + \lambda_i x\right),
    \label{eq:dify}\\
    \lambda_i' &=& -\alpha h H^{-\alpha}\frac{V_{i.\phi}}{V_i}+H^{-\alpha}\phi'\frac{V_{i,\phi\phi}}{V_i} - H^{-\alpha}\frac{V_{i,\phi}^2}{V_i^2}\phi'=\lambda_i\left(-\alpha h + \lambda_ix(\mu_i-1)\right),
    \label{eq:diflambda}
\end{eqnarray}
where $i=j,p,g,r$ and $\beta_i$ denotes the exponents of $H$ in Eq.~(\ref{eq:ys}). Knowing how these variables evolve, we can easily find their fixed points. Considering constant $\mu_i$ allows for potentials in a power law form. For the $y_i$ and the $\lambda_i$, these fixed points are those where
\begin{eqnarray}
    \beta_i h + \lambda_i x&=&0,
    \label{eq:fixedpointy}\\
   \lambda_i x (\mu_i -1)&=& \alpha h .
   \label{eq:fixedpointlambda}
\end{eqnarray}
In this fixed point, combining Eqs.~(\ref{eq:fixedpointy})-(\ref{eq:fixedpointlambda}) yields
\begin{equation}
    \lambda_i\left(-\frac{\beta_i h}{\lambda_i}\right)(\mu_i -1)= \alpha h \Rightarrow (1-\mu_i)=\frac{\alpha}{\beta_i},
\end{equation}
which allows us to directly find the exponents for all four potentials, and they become
\begin{eqnarray}
    V_j(\phi)&=&c_j\phi^{\frac{4}{\alpha}-2},
    \label{eq:potentialj}\\
    V_p(\phi)&=&c_p\phi^{\frac{6}{\alpha}-3},
    \label{eq:potentialp}\\
    V_g(\phi)&=&c_g\phi^{\frac{2}{\alpha}},
    \label{eq:potentialg}\\
    V_{r,\phi}(\phi)&=&c_r\phi^{\frac{4}{\alpha}-1}.
    \label{eq:potentialr}
\end{eqnarray}
Furthermore, the fixed point for $\sigma$ corresponds to $\sigma(h+1)=0$, i.e. fixed points solutions exist for $\sigma=0$ and $h=-1$. On this line of thought, we are interested in the case where $\sigma=0$, which refers to vanishing spatial curvature, as it is consistent with observations \cite{Planck_2020}. In this selected fixed point, the full scalar field equation  Eq.~(\ref{eq:eomr}) by using also Eqs.~\eqref{eq:fixedpointy}-\eqref{eq:fixedpointlambda} can be written as:
\begin{equation}
    2xy_j(h+3) - 3x^2y_p(h+3)-2\lambda_gy_g(h+2) - 8y_r(h+1)=0.
    \label{eq:eomfixedpoint}
\end{equation}
Similarly, the Hamiltonian constraint Eq.~(\ref{eq:hambuena}) can be written as
\begin{equation}
    9x^2y_j - 15x^3y_p - 6y_g(1-2h)-24xy_r = -\rho_{\Lambda}.
    \label{eq:hamfixedpoint}
\end{equation}
Following the arguments in \cite{Copeland:2012qf}, because the self-tuning filter demands that the solutions hold for any value of the cosmological constant, the parameters in the potentials ($\alpha$, $c_i$) should be all independent of $\rho_{\Lambda}$. This assumption constrains the possible values of the terms present in Eqs.~(\ref{eq:eomfixedpoint})-(\ref{eq:hamfixedpoint}) while satisfying the following 
\begin{eqnarray}
    (3+h)c_p(-\alpha h)^{2-\frac{2}{\alpha}}&=&0, \label{eq:constraint1}\\
    -2c_j\alpha h (3+h) - 8c_r(h+1)&=&0, \label{eq:constraint2}\\
    h(2+h)c_g(-\alpha h)^{-1+\frac{2}{\alpha}}&=&0. \label{eq:constraint3}
\end{eqnarray}
We can distinguish several values for $h$ that fulfill these requirements, which are associated with different behaviors of $a$, due to $H=H_0 a^h$. Two cases of interest in this article are arbitrary for $h\neq 0$ and radiation for $h=-2$.

Under the arbitrary case, we focus on three cases:
on the case $h=-1$, which corresponds to an evolution that goes like $H^2\propto a^{-2}$, that is, an accelerated expansion; the case $h=-3/2$, with evolution like $H^2\propto a^{-3}$, corresponding to matter-like expansion. And finally, the radiation case, due to their applicability in studying the standard cosmological model, which corresponds to an inflationary period followed by a radiation-dominated epoch, then a matter-like expansion where structure formation happens, and finally followed by a late-time inflationary expansion that we are currently observing \cite{dodelson_schmidt_2021}. The applicability of such cases is due to the fact that we are looking for a viable cosmological scheme that can reproduce the current cosmic dynamics. Of course, in this case, we are considering solely a particular scalar field dynamic minimally coupled to gravity, so, the possibility that this scalar field can imitate dark matter and the effects of dark energy, could set up a \textit{toy} contemporary universe model.

%%%%%%%%%%%%%%%%%%%%%%%%%%%%%%%%%%%%%%%%%%%%
%%%%%%%%%%%%%%%%%%%%%%%%%%%%%%%%%%%%%%%%%%%%
\section{Fab Four cosmography}
\label{sec:fabfour_cosmography}

In what follows, we proceed with the standard cosmographic methodology \cite{Bamba:2012cp,Capozziello:2019cav,Escamilla-Rivera:2019aol} to derive the kinematic parameters, which will be evaluated at current cosmic times. The advantage to consider this approach is the geometry construction behind it. By assuming the structure of the Cosmological Principle we can factorize the scale factor around the present time. However, we should take into account that we require to relax the issues of such approach: (i) the convergence of the series, which can be attended by using observables in order to restrain the order of it. (ii) the cosmic systematics that could be propagated on the cosmographic parameters. In such case, we require a control of the numerical fit.
Notice that for this analysis we require to select a fixed stable point and two possible cosmological scenarios.

Staying in the fixed point $\sigma=0$, we can start by constructing cosmographic series based on the Taylor series expansion context of the scale factor $a$, defined as
\begin{equation}
    a = 1+ H_0\Delta t - \frac{1}{2}q_0H_0^2\Delta t^2 + \mathcal{O}(\Delta t^3),
    \label{eq:cosmographical}
\end{equation}
where $q$ defined as
\begin{equation}
    q=-\frac{\ddot{a} a}{\dot{a}^2} = -\frac{\dot{H}}{H^2}-1=-\frac{H'}{H}-1   ,
    \label{eq:deceleration}
\end{equation}
and $q_0$ is the current deceleration parameter.
To obtain $H_0$, we first choose an appropriate scenario as described in Sec.~\ref{sec:fabfour-intro}, substitute it into the Hamiltonian constraint Eq.~(\ref{eq:hambuena}), and solve the system for the Hubble parameter $H$. Furthermore, with $H$ we can find $q$ using Eq.~(\ref{eq:deceleration}). Finally, we can then evaluate both expressions at the current time to obtain $H_0$ and $q_0$, respectively. 

Specifically, we chose to build the cosmographical series because it is a purely kinematic expression where other parameters such as matter density or curvature do not play any role. Therefore, we can focus solely on the kinematic terms without specifying other parameters of the model.

At this point, we consider the radiation-dominant behaviour case (for the early-universe calculation of $H_0$), and the arbitrary case (for the late-universe calculation of $H_0$). 
On one hand, for the arbitrary case, we have only two potential functions different from zero (as can be seen in Eqs. \eqref{eq:constraint1}-\eqref{eq:constraint3} by letting $h$ be arbitrary):
\begin{eqnarray}
    V_j(\phi) &=& c_{\text{arb}}(h+1)\phi^{\frac{4}{\alpha}-2},\\
    V_{r,\phi}(\phi) &=& -\frac{\alpha}{4}h(3+h)c_{\text{arb}}\phi^{\frac{4}{\alpha}-1},
\end{eqnarray}
where the subindex \textit{arb} denotes arbitrary.
Therefore, the only Hamiltonian parts considered would be $\ham_j$ and $\ham_r$, thus we can write Eq.~(\ref{eq:hambuena}) as
\begin{eqnarray}
9c_{\text{arb}}(1+h)H^4{\phi'}^2\phi^{\frac{4}{\alpha}-2}+6c_{\text{arb}}\alpha h(3+h)H^4\phi'\phi^{\frac{4}{\alpha}-1}+\rho_\Lambda &=&0, \\
-3c_{\text{arb}}H^4\phi' \phi^{\frac{4}{\alpha}-1}\left[3(h+1)\phi'\phi^{-1} + 2\alpha h(3+h)\right]&=&\rho_\Lambda.
    \label{eq:hamarbitrario}
\end{eqnarray}
From the latter we can solve for $H$:
\begin{equation}
    H^{\text{arb}}=\frac{\sqrt[4]{\rho_\Lambda}\phi^{-\frac{1}{\alpha}+\frac{1}{4}}}{\sqrt[4]{3c_{\text{arb}}\phi'}\sqrt[4]{-3(h+1)\phi'\phi^{-1} - 2\alpha h(3+h)}},
    \label{eq:harb}
\end{equation}
which evaluating at current times gives
\begin{equation}
    H_0^{\text{arb}}=\frac{\sqrt[4]{\rho_{\Lambda 0}}\phi_0^{-\frac{1}{\alpha}+\frac{1}{4}}}{\sqrt[4]{3c_{\text{arb}}\phi'_0}\sqrt[4]{-3(h+1)\phi'_0 \phi_0^{-1} - 2\alpha h(3+h)}}
    \label{eq:h0arb},
\end{equation}
where we have discarded the negative signs of the fourth roots to ensure a real, positive value of $H_0$.
Substituting Eq.~(\ref{eq:harb}) in Eq.~(\ref{eq:deceleration}), and evaluating at the current time gives us
\begin{align}
q_0^{\text{arb}} = -1-\frac{\rho'_{\Lambda 0}}{4\rho_{\Lambda 0}}- \frac{h\alpha\phi_0{\phi_0'}^2(3+h)(\alpha-4)+3{\phi_0'}^3(h+1)(\alpha-2)-h\alpha^2\phi_0^2\phi_0''(3+h)}{2\alpha\phi_0\phi'_0\left[2h\alpha \phi_0(3+h) + 3\phi'_0(h+1)\right]}\nonumber\\
    -\frac{3\alpha\phi_0 \phi'_0 \phi''_0(h+1)}{2\alpha\phi_0\phi'_0\left[2h\alpha \phi_0(3+h) + 3\phi'_0(h+1)\right]}.
    \label{eq:q0arb}
\end{align}

On the other hand, for the radiation scenario, the potentials to be considered are 
\begin{eqnarray}
    V_j(\phi) &=& c_1\phi^{\frac{4}{\alpha}-2},\\
    V_g(\phi) &=& c_2\phi^{\frac{2}{\alpha}},\\
    V_{r,\phi}(\phi) &=& -\frac{\alpha}{2}c_1\phi^{\frac{4}{\alpha}-1},
\end{eqnarray}
which come from considering $h=-2$ in Eqs. \eqref{eq:constraint1}-\eqref{eq:constraint3}; with the corresponding Hamiltonian parts $\ham_j$, $\ham_g$ and $\ham_r$. The full Hamiltonian constraint for this case is
\begin{eqnarray}
    9c_1H^4{\phi'}^2\phi^{\frac{4}{\alpha}-2}-6c_2 H^2 \phi^{\frac{2}{\alpha}}\left(1+\frac{2}{\alpha}\phi^{-1}\phi' \right) +12c_1\alpha H^4\phi'\phi^{\frac{4}{\alpha}-1}+\rho_\Lambda =0,
    \label{eq:hamradiation}
\end{eqnarray}
where solving for $H$ we obtain
\begin{equation}
    H^{\text{rad}}_{\pm}= \sqrt{\frac{\alpha \rho_{\Lambda} \phi^{1-\frac{2}{\alpha}}}{ 3c_2(\alpha\phi + 2\phi') \pm \alpha \gamma(\alpha)}}
    \label{eq:hrad}.
\end{equation}
Here, we have defined the quantity
\begin{eqnarray}
    \gamma(\alpha) =  \frac{1}{\alpha}\sqrt{9c_2^2\alpha^2\phi^2-12\alpha\phi\phi'(-3c_2^2+c_1\alpha^2\rho_\Lambda) + 9{\phi'}^2(4c_2^2-c_1\alpha^2\rho_\Lambda)},
    \label{eq:gamma}
\end{eqnarray}
which will appear frequently in our results.
From Eq.~(\ref{eq:hrad}) and using Eq.~(\ref{eq:gamma}) we can obtain values for $H_0$ and $q_0$ for the radiation scenario:
\begin{eqnarray}
H_{0\pm}^{\text{rad}}&=&\sqrt{\frac{\alpha \rho_{\Lambda 0} \phi_0^{1-\frac{2}{\alpha}}}{ 3c_2(\alpha\phi_0 + 2\phi'_0) \pm \alpha \gamma_0(\alpha)}},
    \label{eq:h0rad}\\
    q_0^{\text{rad}}&=&-1 +\frac{1}{2} \left[ -\frac{\rho_{\Lambda 0}'}{\rho_{\Lambda 0}} -\frac{(\alpha-2)\phi_0'}{\alpha\phi_0}
   +\frac{3\alpha\phi_0' [ 2c_2 (3\alpha c_2\phi_0 +6c_2\phi_0'\pm\alpha\gamma) ] %\right.
   }{2\alpha\gamma ( 3\alpha c_2\phi_0+6c_2\phi_0' \pm \alpha\gamma )}
    \right.
    \nonumber \\ 
  %  &&
   && \left.
    +\frac{6\phi_0'' [ 2c_2 ( 3\alpha c_2\phi_0+6c_2\phi_0'\pm\alpha\gamma )-\alpha^2c_1\rho_{\Lambda 0} ( 2\alpha\phi_0+3\phi_0' ) ]}{2\alpha\gamma ( 3\alpha c_2\phi_0+6c_2\phi_0' \pm \alpha\gamma )} \right]. \nonumber \\
&& \left.-\frac{3\alpha^2 c_1\phi_0' ( \rho_{\Lambda 0}' (4\alpha\phi_0+3 \phi_0' )+4\alpha\rho_{\Lambda 0}\phi_0'  )}{2\alpha\gamma ( 3\alpha c_2\phi_0+6c_2\phi_0' \pm \alpha\gamma )} \right].    \label{eq:q0rad}
\end{eqnarray}

Now, in the calculated values for $H_0$ and $q_0$ in Eqs.~(\ref{eq:h0arb})-(\ref{eq:q0arb}), (\ref{eq:h0rad})-(\ref{eq:q0rad}) there are several free parameters, namely: $c_{\text{arb}}$, $c_1$, $c_2$, $\alpha$, $h$ and the scalar field and its derivative at current time $\phi_0$ and $\phi_0'$. 
We can choose an arbitrary value of $\alpha$ to fix the freedom in re-defining the scalar field. Following \cite{Copeland:2012qf}, we can select $\alpha$ such that: $\alpha h =-1$. Alternatively, we can choose values of $\alpha$ to ensure we can obtain power law potentials of the form $V\propto\phi^2$, which gives an inflationary behaviour.  Only the cases where $\alpha=1$ and $\alpha=2$, give potentials of this kind, while $\alpha=1$ is preferred as it also fulfills $\alpha h =-1$ when $h=-1$ (inflation).

%%%%%%%%%%%%%%%%%%%%%%%%%%%%%%%%%%%%%%%%%%%%
%%%%%%%%%%%%%%%%%%%%%%%%%%%%%%%%%%%%%%%%%%%%

\subsection{Case for $\alpha=1$}

For this case, we consider such $\alpha=1$ value in Eqs.~(\ref{eq:h0arb})-(\ref{eq:q0arb}), which
yields
\begin{equation}
    H_0^{\text{arb}}=\frac{\sqrt[4]{\rho_{\Lambda 0}}}{\phi_0^{\frac{3}{4}}\sqrt[4]{3c_{\text{arb}}\phi'_0}\sqrt[4]{-3(h+1)\phi'_0 \phi_0^{-1} - 2 h(3+h)}},
    \label{eq:h0arbalfa1}
\end{equation}
and
\begin{equation}
    q_0^{\text{arb}} = -1-\frac{\rho'_{\Lambda 0}}{4\rho_{\Lambda 0}}+ \frac{3h\phi_0{\phi_0'}^2(3+h)+3{\phi_0'}^3(h+1)+h\phi_0^2\phi_0''(3+h)+3\phi_0 \phi'_0 \phi''_0(h+1)}{2\phi_0\phi'_0\left[2h\phi_0(3+h) + 3\phi'_0(h+1)\right]}.
    \label{eq:q0arbalfa1}
\end{equation}
Now, choosing a value for $h$ determines how the Hubble parameter depends on $a$ according to Eq.~(\ref{eq:hysigma}), and thus specifies the type of expansion. We can study the cases for $h=-3/2$ and $h=-1$, which correspond to matter-dominated expansion and inflation, respectively.
Substituting these values of $h$ in Eqs.~(\ref{eq:h0arbalfa1})-(\ref{eq:q0arbalfa1}) gives
\begin{equation}
    H_0^{\text{matter}} = \frac{\sqrt[4]{2\rho_{\Lambda 0}}}{\sqrt{3}\phi_0^{\frac{3}{4}}\sqrt[4]{c_{\text{arb}}\phi'_0\left(3+\phi'_0 \phi_0^{-1}\right)}},
    \label{eq:h0matteralfa1}
\end{equation}
\begin{eqnarray}
    q_0^{\text{matter}} = -1-\frac{\rho_{\Lambda 0}'}{4\rho_{\Lambda 0}} + \frac{{\phi_0'}^2(9\phi_0 + 2\phi_0')+\phi_0 \phi_0''(3\phi_0 + 2\phi_0')}{4\phi_0\phi_0'(3\phi_0 + \phi_0')},
    \label{eq:q0matteralfa1}
\end{eqnarray}
which correspond to $h=-3/2$, and
\begin{eqnarray}
    H_0^{\text{inf}}&=&\frac{\sqrt[4]{\rho_{\Lambda 0}}}{\sqrt{2}\phi_0^{\frac{3}{4}}\sqrt[4]{3c_{arb}\phi_0'}},
    \label{eq:h0infalfa1}\\
    q_0^{\text{inf}}&=&-1-\frac{\rho_{\Lambda 0}'}{4\rho_{\Lambda 0}} + \frac{3\phi_0'}{4\phi_0} + \frac{\phi_0''}{4\phi_0'},
    \label{eq:q0infalfa1}
\end{eqnarray}
which corresponds to the inflationary case $h=-1$.

Now, the case for radiation-like expansion is given by substituting $\alpha=1$ in $\gamma(\alpha)$ as in Eqs.~(\ref{eq:h0rad})-(\ref{eq:q0rad}):
\begin{equation}
    H_{0 \pm}^{\text{rad}}=\sqrt{\frac{\rho_{\Lambda 0 }\phi_0^{-1}}{3c_2(\phi_0 + 2\phi_0') \pm \gamma_0(1)}},
    \label{eq:h0radalfa1}
\end{equation}
\begin{eqnarray}
    q_{0\pm}^{\text{rad}}&=&-1-\frac{\rho_{\Lambda 0}'}{2\rho_{\Lambda 0}} + \frac{\phi_0'}{2\phi_0} + \frac{3\phi_0'\left(2c_2(3c_2\phi_0 + 6c_2\phi_0' \pm \gamma(1)) - c_1(\rho_{\Lambda 0}'(4\phi_0 + 3\phi_0') + 4\rho_{\Lambda 0}\phi_0')\right)}{4\gamma(1)\left(3c_2\phi_0 + 6c_2\phi_0' \pm \gamma(1)\right)} \nonumber \\
  &&  + \frac{6\phi_0''\left(2c_2(3c_2\phi_0 + 6c_2\phi_0' \pm \gamma(1))-c_1\rho_{\Lambda 0}(2\phi_0+3\phi_0') \right)}{4\gamma(1)\left(3c_2\phi_0 + 6c_2\phi_0' \pm \gamma(1)\right)}
    \label{eq:q0radalfa1}.
\end{eqnarray}

%%%%%%%%%%%%%%%%%%%%%%%%%%%%%%%%%%%%%%%%%%%%
%%%%%%%%%%%%%%%%%%%%%%%%%%%%%%%%%%%%%%%%%%%%

\subsection{Case for $\alpha=2$}

Another way to obtain potentials of the form $\phi^2$ is with $\alpha=2$. Substituting this value on Eqs.(\ref{eq:h0arb})-(\ref{eq:q0arb}),~(\ref{eq:h0rad})-(\ref{eq:q0rad}), and following a similar procedure as in the latter case, we obtain
\begin{equation}
    H_0^{\text{arb}} = \frac{\sqrt[4]{\rho_{\Lambda 0}}}{\sqrt[4]{3c_{\text{arb}}\phi_0\phi_0'}\sqrt[4]{-3(h+1)\phi_0'\phi_0^{-1} - 4h(3+h)}},
    \label{eq:h0arbalfa2}
\end{equation}
\begin{equation}
    q_0^{\text{arb}}=-1-\frac{\rho_{\Lambda 0}'}{4\rho_{\Lambda 0}} + \frac{2h(h+3)({\phi_0'}^2 + \phi_0 \phi_0'') + 3(h+1)\phi_0' \phi_0''}{2\phi_0'(4h\phi_0(3+h) + 3\phi_0'(h+1))},
    \label{q0arbalfa2}
\end{equation}
and for the matter case
\begin{eqnarray}
    H_0^{\text{matter}}&=&\frac{\sqrt[4]{\rho_{\Lambda 0}}}{\sqrt[4]{3c_{\text{arb}}\phi_0 \phi_0'}\sqrt[4]{9+\frac{3\phi_0'}{2\phi_0}}},
    \label{eq:h0matteralfa2}\\
    q_0^{\text{matter}}&=&-1-\frac{\rho_{\Lambda 0}'}{4\rho_{\Lambda 0}} + \frac{3{\phi_0'}^2 + \phi_0''(\phi_0+\phi_0')}{2\phi_0'(6\phi_0 + \phi_0')}.
    \label{eq:q0matteralfa2}
\end{eqnarray}
The inflation case gives
\begin{eqnarray}
    H_0^{\text{inf}}&=&\frac{\sqrt[4]{\rho_{\Lambda 0}}}{2^{\frac{3}{4}}\sqrt[4]{3c_{\text{arb}}\phi_0\phi_0'}},
    \label{eq:h0infalfa2}\\
    q_0^{\text{inf}}&=&-1-\frac{\rho_{\Lambda 0}'}{4\rho_{\Lambda 0}}+\frac{\phi_0'}{4\phi_0} + \frac{\phi_0''}{4\phi_0'},
    \label{eq:q0infalfa2}
\end{eqnarray}
\begin{equation}
    H_{0\pm}^{\text{rad}}=\sqrt{\frac{\rho_{\Lambda 0}}{3c_2(\phi_0+\phi_0')\pm \gamma(2)}},
    \label{eq:h0radalfa2}
\end{equation}
and for the radiation case
\begin{eqnarray}
    q_{0\pm}^{\text{rad}}=-1-\frac{\rho_{\Lambda 0}'}{2\rho_{\Lambda 0}} + \frac{\pm6c_2\phi_0' \gamma(2) \pm 6c_2\phi_0''\gamma(2) + 18c_2^2(\phi_0+\phi_0')(\phi_0'+  \phi_0'')}{4\gamma(2)(3c_2(\phi_0 + \phi_0') + \gamma(2))}\nonumber \\- \frac{3c_1(8\rho_{\Lambda 0}{\phi_0'}^2 + \rho_{\Lambda 0}'\phi_0'(8\phi_0 + 3\phi_0') + 2\rho_{\Lambda 0}\phi_0''(4\phi_0 + 3\phi_0'))}{4\gamma(2)(3c_2(\phi_0 + \phi_0') + \gamma(2))}.
    \label{eq:q0radalfa2}
\end{eqnarray}
For both values of $\alpha$, we notice similarities in all forms of $H_0$, which are all proportional to $\propto \sqrt[4]{\rho_\Lambda}$. Additionally, we notice similarities in the expressions for $q_0$, where all of them can be written in the same functional form
\begin{equation}
    q_0 = -1 - A \left(\frac{\rho'_{\Lambda}}{\rho_\Lambda} \right)+ B \left(\frac{\phi'}{\phi} \right)+ C \left(\frac{\phi''}{\phi'}\right),
    \label{eq:q0general}
\end{equation}
where $A$, $B$ and $C$ are, in general, functions of the parameters $\rho_\Lambda$, $\phi$, $\phi'$, $\phi''$ and the constants $c_{\text{arb}}$, $c_1$ and $c_2$. Considering standard GR with a cosmological constant and a scalar field as the source in an FLRW background, the deceleration parameter has the same functional form.

%%%%%%%%%%%%%%%%%%%%%%%%%%%%%%%%%%%%%%%%%%%%
%%%%%%%%%%%%%%%%%%%%%%%%%%%%%%%%%%%%%%%%%%%%

\section{Fab-Four on the $H_0$ tension}
\label{sec:fabfour_tension}

According to the latter results, we can now discuss the possible requirements to relax the value of $H_0$ in the cosmological scenarios discussed.

With the expressions obtained for $H_0^{\text{inf}}$ and $H_0^{\text{rad}}$ we can now calculate the predicted discrepancy between both values. The most recent measurements \cite{Planck_2020, Riess_2022} suggest a discrepancy of around $\sim$ 10\% between the values for the Hubble constant in the early universe and those in the late universe. As such, we expect a
\begin{equation}
    \delta_{H_0} \equiv 1-\frac{H_0^{\text{rad}}}{H_0^{\text{inf}}} \approx 0.1
    \label{eq:deltareal}
\end{equation}
For $\alpha=1$ we can use Eqs.(\ref{eq:h0infalfa1})-(\ref{eq:h0radalfa1}) to get:
\begin{equation}
    \delta_{H_0, \alpha=1} = 1- \frac{\sqrt{2}\sqrt[4]{3c_{\text{arb}}\rho_{\Lambda 0}\phi_0 \phi_0'}}{\sqrt{3c_2(\phi_0 + 2\phi_0')\pm \gamma(1)}}.
    \label{eq:deltaalfa1}
\end{equation}
On the other hand, for $\alpha=2$, using Eqs.~(\ref{eq:h0infalfa2})-(\ref{eq:h0radalfa2}), we obtain the following
\begin{equation}
    \delta_{H_0, \alpha=2} = 1- \frac{\sqrt[4]{24c_{\text{arb}}\rho_{\Lambda 0}\phi_0 \phi_0'}}{\sqrt{3c_2(\phi_0 + \phi_0')\pm \gamma(2)}}
    \label{eq:deltaalfa2}.
\end{equation}
Now, before continuing, throughout the procedure to obtain expressions for $H_0$, $q_0$, and $\delta_{H_0}$, several requirements emerge so the solutions are real. These are summarized in Table \ref{tab:requirements}.

\begin{table*}[t]
    \centering
    \begin{tabular}{|c|c|c|}
        \hline
         \multicolumn{2}{|c|}{\textbf{Case}} & \textbf{Requirements} \\ \hline
         \multicolumn{2}{|c|}{General} & $\rho_{\Lambda 0}>0, \qquad c_{arb}\phi'_0>0, \qquad -3(h+1)\phi'_0 \phi_0^{-1} - 2\alpha h(3+h)>0, \qquad \phi_0 >0$ \\ \hline
         \multirow{4}{*}{$\alpha=1$} & \multirow{2}{*}{General} & $9c_2^2\phi_0^2-12\phi_0\phi_0'(-3c_2^2+c_1\rho_{\Lambda 0}) + 9{\phi_0'}^2(4c_2^2-c_1\rho_{\Lambda 0})>0$, \\
          &  & $3c_2(\phi_0 + 2\phi_0') \pm \gamma_0(1)>0$\\ \cline{2-3}
          & $h=-3/2$ (Matter) & $3+\phi'_0\phi^{-1}_0>0$ \\ \cline{2-3}
          & $h=-1$ (Inflation) & No extra requirements \\ \hline
         \multirow{4}{*}{$\alpha=2$} & \multirow{2}{*}{General} & $9c_2^2\phi_0^2 + 6\phi_0 \phi_0'(3c_2^2-4c_1\rho_{\Lambda 0}) + 9{\phi_0'}^2(c_2^2-c_1\rho_{\Lambda 0})>0$ \\
         & & $3c_2(\phi_0 + \phi_0') \pm \gamma_0(2) >0$\\ \cline{2-3}
         & $h=-3/2$ (Matter) & $6+\phi_0' \phi_0^{-1}>0$ \\ \cline{2-3}
         & $h=-1$ (Inflation) & No extra requirements \\ \hline
    \end{tabular}
    \caption{Requirements on the Fab-Four derived constants in the results for $H_0$, $q_0$, and $\delta_{H_0}$.}
    \label{tab:requirements}
\end{table*}

Taking into account the restrictions given by Table \ref{tab:requirements}, it is possible to get values for $\delta_{H_0}=0.1$, by considering the requirements on the five free parameters that appear in the expressions: $\phi_0, \: \phi'_0, \: c_{\text{arb}}, \: c_1$ and $c_2$. We also have another two parameters, $\rho'_\Lambda$ and $\phi''$, which only appear in the expressions for $q_0$. 

However, the behaviour of $\rho_{\Lambda}$ comes into play. The Fab-Four model was introduced to handle the cosmological constant fine-tuning problem \cite{Charmousis_2012,Copeland:2012qf}. As such, instead of the cosmological constant being constant, it is assumed that it is a piecewise constant function, with several phase transitions that change its value. Therefore, $\rho_{\Lambda}$ is discontinuous at these phase transitions. This means that the \textit{r.h.s} of Eq.~(\ref{eq:hambuena}) is discontinuous, but the \textit{l.h.s} of this equation is composed of the Hamiltonian parts, which depend only on the functions $a$, $\phi$, and $\phi'$. Since $a$ is continuous, and $\phi$ as well - by construction - it follows that $\phi'$ must be discontinuous to maintain equality at the phase transitions. Consequently, both $\rho'_\Lambda$ and $\phi''$ contain delta-functions localized at the transition times. These same arguments are used to initially restrict Horndeski's full theory of gravity \cite{Charmousis_2012}.

At these phase-transitions, the self-tuning nature of the Fab-Four theory is active to still allow a standard cosmology to develop \cite{Charmousis_2012}. However, by construction, the self-tuning filter can only work when the curvature is non-zero. Since the latest observations assert a negligible curvature \cite{Planck_2020}, we assume that we are not presently in a phase transition. 
For this work, we are only taking the present-day values of these parameters ($\rho'_{\Lambda 0}$, $\phi''_0$), which we will assume to be zero since we are not currently in a phase-transition, where there would be a delta-function. \\

Therefore, substituting $\rho'_{\Lambda 0}=0$ and $\phi''_0 =0$, from Eqs.~(\ref{eq:q0infalfa1})-(\ref{eq:q0radalfa1}), (\ref{eq:q0infalfa2})-(\ref{eq:q0radalfa2}) we obtain
\begin{eqnarray}
    q^{\text{inf}}_{0 \; \alpha=1}&=&-1 + \frac{3\phi_0'}{4\phi_0},
    \label{eq:q0infalfa1simple}\\
    q_{0 \; \alpha=1}^{\text{rad}}&=&-1 + \frac{\phi_0'}{2\phi_0} + \frac{3\phi_0'\left(2c_2(3c_2\phi_0 + 6c_2\phi_0' \pm \gamma(1)) - 4c_1\rho_{\Lambda 0}\phi_0'\right)}{4\gamma(1)\left(3c_2\phi_0 + 6c_2\phi_0' \pm \gamma(1)\right)},
    \label{eq:q0radalfa1simple}\\
    q^{\text{inf}}_{0 \; \alpha=2}&=&-1+\frac{\phi_0'}{4\phi_0}
    \label{eq:q0infalfa2simple},\\
    q_{0 \; \alpha=2}^{\text{rad}}&=&-1 + \frac{\pm3c_2\phi_0' \gamma(2) + 9c_2^2\phi_0'(\phi_0+\phi_0')-12c_1\rho_{\Lambda 0}{\phi_0'}^2}{2\gamma(2)(3c_2(\phi_0 + \phi_0') + \gamma(2))}
    \label{eq:q0radalfa2simple},
\end{eqnarray}
for each $\alpha=1,2$ case, respectively. Now, given a value of $q_0$ through the latest observations \cite{Riess_2022} ($q_0\approx -0.5$), we can either force the deceleration parameters (evaluated at the current time) of both the early and late universe to be the same, or we can leave them to be different. 

In this work, we assume $q_0^{\text{inf}}=q_0^{\text{rad}}\approx -0.5$ for simplicity, and we leave the other case for future work. Given this restriction, with the expressions for $q_0^{\text{inf}}$ Eqs.~(\ref{eq:q0infalfa1simple}),(\ref{eq:q0infalfa2simple}), we will get a fixed value for $\phi'_0 / \phi_0$ (depending on the choice of $\alpha$). Then using the expressions for $q_0^{\text{rad}}$ Eqs.~(\ref{eq:q0radalfa1simple}),(\ref{eq:q0radalfa2simple}) we can constrain the value of one more parameter, say for example, $c_1$.

Finally, we are only left with two free parameters: $c_{\text{arb}}$ and either $c_1$ or $c_2$. These are enough to get $\delta_{H_0}\approx0.1$, and reducing the number of free parameters is important to enhance the likelihood of this model in Bayesian statistic comparisons \cite{escamilla2023}.

%%%%%%%%%%%%%%%%%%%%%%%%%%%%%%%%%%%%%%%%%%%%
%%%%%%%%%%%%%%%%%%%%%%%%%%%%%%%%%%%%%%%%%%%%

\section{Conclusions} 
\label{sec:conclusions}
 
In this work, we consider the possibility to tackle the Hubble tension using the Fab-Four theory. This scalar-tensor theory of gravity applies a self-tuning filter on the more general Horndeski theory, to solve the cosmological constant fine-tuning problem. This leaves a much more manageable theory, consisting only of four base Lagrangians given by Eqs.\eqref{eq:lagjconmetrica}-\eqref{eq:lagrconmetrica}. We follow the procedure discussed in \cite{Copeland:2012qf} to obtain a simpler form of the theory, using a dynamical system approach to find the fixed points of the theory, and we concentrate on the $\sigma=0$ fixed-point, which denotes zero curvature (as supported by observations \cite{Planck_2020}). In this fixed point, the equation of motion of the scalar field Eq.~\eqref{eq:eomfixedpoint} and the Hamiltonian restriction Eq.~\eqref{eq:hamfixedpoint} of the theory are such that we have certain allowed types of cosmological scenarios, among which are the inflationary expansion, radiation-like expansion, and a matter-like phase. This shows the utility of the Fab-Four theory, as \textit{turning on-and-off} certain potentials lead naturally to different types of cosmological behaviors, and the full dynamics of the theory allow for a smooth, continuous shift between these types of cosmological evolution.

Furthermore, we construct the Fab-Four cosmography in the zero-curvature fixed point. In particular, we derive analytical expressions for the current Hubble constant $H_0$, and the deceleration parameter $q_0$, for the matter, inflation, and radiation scenarios, as well as for different choices of the arbitrary constant $\alpha$. To do this, we take into account the forms of the potentials, substitute them into the Hamiltonian constraint, which results in quartic equations of the Hubble parameter, and then solve for $H$. These equations yield four solutions: \textit{(i)} in the arbitrary case, we discard two complex roots and a negative one; \textit{(ii)} in the radiation case, we can only discard two negative roots, leaving both signs of the expression for $H^{\text{rad}}_\pm$. Examining Table \ref{tab:requirements}, we also notice that the requirements do not restrict the choice of signs in the expressions for $H^{\text{rad}}_\pm$ and $q^{\text{rad}}_\pm$, and both signs can give $\delta_{H_0}=0.1$,  which is why we have left them written explicitly in all expressions.
Afterward, we differentiate the results for $H$ according to Eq.~\eqref{eq:deceleration} to obtain expressions for the deceleration parameter, which all ultimately take the functional form of Eq.~\eqref{eq:q0general}.  Using these expressions evaluated at current time, we can compare the Hubble constant in the late universe with the Hubble constant in the early universe (radiation-like expansion). 

Finally, we show that it is possible to get the desired discrepancy between the early-universe and the late-universe Hubble constants, of around $\sim 10$\%, showing that the Hubble constant could, in reality, be dynamical. Therefore, we have succeeded in our objective to show that it is indeed possible to get a disagreement between the observed Hubble constant if measured through local means, and the one obtained from observations of the CMB, due to the different dynamics put forward by the Fab-Four theory.

The final results contain five arbitrary parameters which are constrained by the requirements in Table \ref{tab:requirements}. However, using the expressions for the deceleration parameter we can reduce the number of free parameters in these results to only two constants. In future work, we will propose performing an analysis with observational data to obtain confidence limits on the free variables of the theory. For this, extracting other observables from the Fab Four theory, e.g., expressions for the luminosity distance, and even perturbation theory results such as the angular power spectrum from the CMB would be convenient.  Afterward, a possible course of action would be to make a Bayesian statistics comparison of this model's prediction against the standard $\Lambda$CDM to test its plausibility, as well as to further constrain the possible values of the free parameters to match observations \cite{escamilla2023}. Furthermore, the possibility to explore beyond Horndeski theories \cite{Kobayashi:2019hrl} could be a path in order to tackle the cosmological tensions at early cosmic times. 

%%%%%%%%%%%%%%%%%%%%%%%%%%%%%%%%%%%%%%%%%%%%
%%%%%%%%%%%%%%%%%%%%%%%%%%%%%%%%%%%%%%%%%%%%

\begin{acknowledgments}
CE-R acknowledges the Royal Astronomical Society as FRAS 10147 and is supported by PAPIIT UNAM Project TA100122. 
SN acknowledges financial support from SEP–CONACYT postgraduate grants program.
This article is based upon work from COST Action CA21136 Addressing observational tensions in cosmology with systematics and fundamental physics (CosmoVerse) supported by COST (European Cooperation in Science and Technology). 
\end{acknowledgments}

%%%%%%%%%%%%%%%%%%%%%%%%%%%%%%%%%%%%%%%%%%%%
%%%%%%%%%%%%%%%%%%%%%%%%%%%%%%%%%%%%%%%%%%%%
\nocite{*}
\bibliographystyle{unsrt}
\bibliography{references}
\end{document}